\documentclass[pr,twocolumn,showpacs,amsmath,nobibnotes,nofootinbib,floatfix]{revtex4}
\usepackage[latin1]{inputenc}
\usepackage{color}
\usepackage{ulem}
\usepackage{graphicx}
\usepackage{bm}
\usepackage{amssymb}
\usepackage{amsmath}

\begin{document}
\title{Gyrotropy and magneto-spatial dispersion effects at intersubband transitions in quantum wells}

\author{L.~E.~Golub}
\email{golub@coherent.ioffe.ru}
\affiliation{Ioffe
Physical-Technical Institute of the Russian Academy of Sciences, 194021
St.~Petersburg, Russia}

\begin{abstract}
Gyrotropic properties of multiple quantum well structures are studied theoretically. 
Symmetry analysis is performed yielding the gyrotropy tensor components for structures grown along [001], [110] and [311] crystallographic directions.
Angular dependences of circular dichroism and natural optical activity signals are established.
Phenomenological model and microscopic theory based on spin-orbit splitting of size-quantized subbands are developed for photon energies close to the energy of the intersubband optical transition. 
Magneto-spatial dispersion effects arising from the diamagnetic shift of the intersubband energy gap linear in the electron momentum are also considered. 
It is demonstrated that the spectral dependence of the gyrotropy and magneto-spatial dispersion constants represents an asymmetrical peak with a degree of asymmetry governed by the mean electron energy. The estimates show that the considered effects are detectable in experiments.
\end{abstract}



\maketitle

\textbf{1. Introduction.}
Gyrotropic systems are defined as media with a spatial dispersion of the first order which is described by a contribution to the dielectric permittivity linear in the light wavevector $\bm q$~\cite{LL8,Kizel}: 
\begin{equation}
	\delta\varepsilon_{\lambda\mu}=	{\rm i} \gamma_{\lambda\mu\nu} q_\nu.
\end{equation}
Here $\bm \gamma(\omega)$ is a third rank tensor antisymmetric with respect to the first two indices. Manifestation of gyrotropy in light transmission experiments are natural optical activity (rotation of the plane of linear polarization) and circular dichroism (helicity-dependent absorption). In systems of $C_{\rm 3v}$, $C_{\rm 4v}$ and $C_{\rm 6v}$ symmetry these effects are forbidden, therefore such systems are called ``weakly gyrotropic''. In these structures one can detect changes of polarization state in measurements of light reflection.

Gyrotropy of bulk semiconductors has been studied in tellurium under interband excitation~\cite{IP74,Farbshtein} and in weakly gyrotropic A$_2$B$_6$ semiconductors of wurtzite type as CdS near excitonic resonances~\cite{IvchSelk}. A$_3$B$_5$ semiconductors having $T_{\rm d}$ point symmetry are non-gyrotropic. Gyrotropy in bulk GaAs can be induced by application of deformation lowering symmetry  to $D_{\rm 2d}$~\cite{Cardona} or at a surface of a semiconductor~\cite{Alperovich}.
The situation is different for two-dimensional semiconductors because all quantum well (QW) structures are gyrotropic. However, the gyrotropy of QWs was probed only in photogalvanic experiments~\cite{SG_EL} while all-optical study of this phenomenon is absent up to date.

Microscopically, gyrotropy of multiple QW structures (MQWs) is caused by spin-orbit splitting of electron energy spectrum.
The splitting is present due to asymmetry of the heteropotential or absence of inversion center in the bulk material. It is described by Rashba and two-dimensional Dresselhaus contributions to the Hamiltonian of spin-orbit interaction given by $\bm k$-linear terms
\begin{equation}
\label{SO}
	\hat{H}_{SO}(\bm k) = \beta_{\mu\nu} \hat{\sigma}_\mu k_\nu,
\end{equation}
where $\bm k$ is a 2D electron wavevector and $\hat{\sigma}_\mu$ are the Pauli matrices.
The gyrotropy tensor $\bm \gamma$ is related to the pseudotensor $\bm \beta$ via
\[
\gamma_{\lambda\mu\nu} \sim e_{\lambda\mu\rho}\beta_{\rho\nu},
\]
where $\bm e$ is the totally antisymmetric third-rank tensor.

Recently optical activity of noncentrosymmetric metals caused by spin-orbit interaction of such kind has been studied theoretically~\cite{OptActMetals} motivated by experimental results~\cite{ExpMetals}. In metals the spin-orbit splitting exceeds far the photon energy, and the effect is inversely proportional to the spin-orbit spitting: $\gamma \sim1/\beta$~\cite{OptActMetals}. In semiconductor MQWs the situation is opposite, and the correction to the dielectric permittivity is linear in the small spin-orbit splitting.

In the presence of a magnetic field $\bm B$, the dielectric permittivity acquires additional terms linear in both $\bm B$ and $\bm q$ which describe magneto-spatial dispersion:
\begin{equation}
	\delta\varepsilon^{(\bm B)}_{\lambda\mu}=	A_{\lambda\mu\rho\nu} B_\rho q_\nu.
\end{equation}
Magneto-spatial dispersion is allowed in all non-centrosymmetric media, therefore it is present in MQWs. In bulk semiconductors this effect has been investigated in both A$_2$B$_6$ and A$_3$B$_5$ materials~\cite{pss,Gogolin}. In MQW structures, the magneto-spatial dispersion is caused by $\bm k$-linear diamagnetic shift of subband dispersions described by the Hamiltonian
\begin{equation}
	\label{H_dia}
	H_{dia} (\bm k) = \alpha_{\mu\nu}B_\mu k_\nu.
\end{equation}

We consider photon energies close to the peak of intersubband absorption in MQWs.

\textbf{2. Symmetry consideration.}
(001)-grown symmetrical QWs have $D_{\rm 2d}$ symmetry with reflection planes perpendicular to the in-plane axes $x \parallel [1\bar{1}0]$ and $y \parallel [110]$. The gyrotropic corrections to the dielectric tensor of MQWs are described by one linearly independent constant:
\begin{equation}
	\delta\varepsilon_{xz}={\rm i}\gamma q_x, \qquad
	\delta\varepsilon_{yz}=-{\rm i}\gamma q_y.
\end{equation}
These terms result in 
circular dichroism: for light propagating in the structure plane, the absorbance has a contribution
\begin{equation}
\label{NOA_D2d}
\eta \propto {\rm Im}\gamma \, P_{\rm circ} \, {q_xq_y\over q},
\end{equation}
where $P_{\rm circ}$ is the circular polarization degree of light. The real part of $\gamma$ determines the natural optical activity signal. Both effects are maximal for light propagating along an in-plane cubic axis.

Asymmetrical (001)-grown QWs have $C_{\rm 2v}$ symmetry with the $C_2$ axis parallel to the growth direction $z$. Gyrotropy in this case is described by two constants:
\begin{equation}
\label{C2v}
	\delta\varepsilon_{xz}={\rm i}\gamma q_x, \qquad
	\delta\varepsilon_{yz}=-{\rm i}\gamma' q_y.
\end{equation}
Change of light polarization in transmission experiments is described by the sum of these constants:
\[
\eta \propto {\rm Im}(\gamma+\gamma') \, P_{\rm circ} {q_xq_y\over q},
\]
while their difference describing weak gyrotropy can be probed in reflection experiments.

Symmetrical (110)-grown MQWs have $C_{\rm 2v}$ symmetry as well, but the $C_2$ axis lies in the QW plane: $C_2 \parallel y$, where $y\parallel [001]$. Therefore the following corrections are present: 
\begin{equation}
\label{110}
	\delta\varepsilon_{xy}=-{\rm i}\gamma_1 q_x, \qquad
	\delta\varepsilon_{yz}=-{\rm i}\gamma_2 q_z.
\end{equation}
Here $z \parallel [110]$ is the growth direction, and $x \parallel [\bar{1}10]$. In transmission experiments on can measure the following angular dependence:
\begin{equation}
\label{NOA_Cs}
\eta \propto {\rm Im}(\gamma_1+\gamma_2) \, P_{\rm circ} {q_xq_z\over q}.
\end{equation}

If the (110)-grown QW is asymmetric, then it belongs to the class $C_{\rm s}$ with only one mirror reflection plane ($yz$). In this case two additional gyrotropic contributions~\eqref{C2v} are allowed by symmetry.

(113)-grown QWs also have $C_{\rm s}$ symmetry therefore the above relations~\eqref{C2v},~\eqref{110} are valid for them if one chooses the axes as $x \parallel [1\bar{1}0]$, $y \parallel [33\bar{2}]$, $z \parallel [113]$.

Magneto-spatial dispersion effects manifest themselves as corrections to $\varepsilon_{zz}$:
\begin{equation}
	\delta\varepsilon^{(\bm B)}_{zz}=	A_{zz\mu\nu} B_\mu q_\nu.
\end{equation}
Real and imaginary parts of $A_{zz\mu\nu}$ result in magnetic field induced birefringence and $\bm B$-dependent absorption, respectively.

In (001) QWs of $C_{\rm 2v}$ symmetry the correction has the form
\begin{equation}
\label{MSD_C_2v}
	\delta\varepsilon^{(\bm B)}_{zz}=	A_1 B_x q_y + A_2 B_y q_x.
\end{equation}
If the MQW structure is symmetric ($D_{\rm 2d}$ point group) then $A_1=A_2$.

For symmetrical (110) QWs we have
\begin{equation}
\label{MSD_110}
	\delta\varepsilon^{(\bm B)}_{zz}=	A_3 B_z q_x + A_4 B_x q_z.
\end{equation}

In (311) MQWs all four constants $A_{1..4}$ are nonzero and different from each other.

\textbf{3. Microscopic model.}
If we ignore the photon momentum ($\bm q=0$), the intersubband absorption is due to direct optical transitions.  Electron wavevectors in the initial state in the ground subband ($\bm k$) and in the final state in the excited subband ($\bm k'$) coincide: $\bm k' = \bm k$. If we also neglect spin-orbit splitting ($\beta_{\mu\nu}=0$), only photons with energy $\hbar \omega = E_{21}$ can be absorbed. Here $E_{21}$ is the energy separation between the ground and the excited subbands, and we assume parabolic energy dispersions $E_{1}(\bm k)=E_2(\bm k) =\hbar^2k^2/(2m)$ with equal effective masses in the subbands. Therefore the spectrum of intersubband absorbance is a sharp peak centered at $E_{21}$: $\eta=\eta_0(\hbar \omega - E_{21})$, where $\eta_0(x)$ is a broadened $\delta$-function.

Account for a final photon momentum makes the optical transitions indirect: $\bm k' = \bm k + \bm q$. Therefore the energy conservation reads:
\begin{equation}
\label{energy}
	\hbar \omega = E_{21} + {\hbar^2\bm k \cdot \bm q \over m} + {\hbar^2 q^2 \over 2m}.
\end{equation}
The presence of the second term here leads to Doppler broadening of the absorption peak~\cite{SG_Prettl_book}, and the third term in Eq.~\eqref{energy} makes a shift of the peak to higher energies. As a result, the absorbance takes the following form:
\begin{equation}
\label{eta_q}
	\eta(\hbar \omega - E_{21})= \eta_0 - {\hbar^2 q^2 \over 2m} \left[{\partial\over\partial (\hbar\omega)} - \overline{E} \: {\partial^2\over\partial (\hbar\omega)^2}\right] \eta_0,
\end{equation}
where $\overline{E}$ is the mean kinetic energy. Hereafter we assume filling of the ground subband only.

\begin{figure*}[tb]
\includegraphics[width=0.3\linewidth]{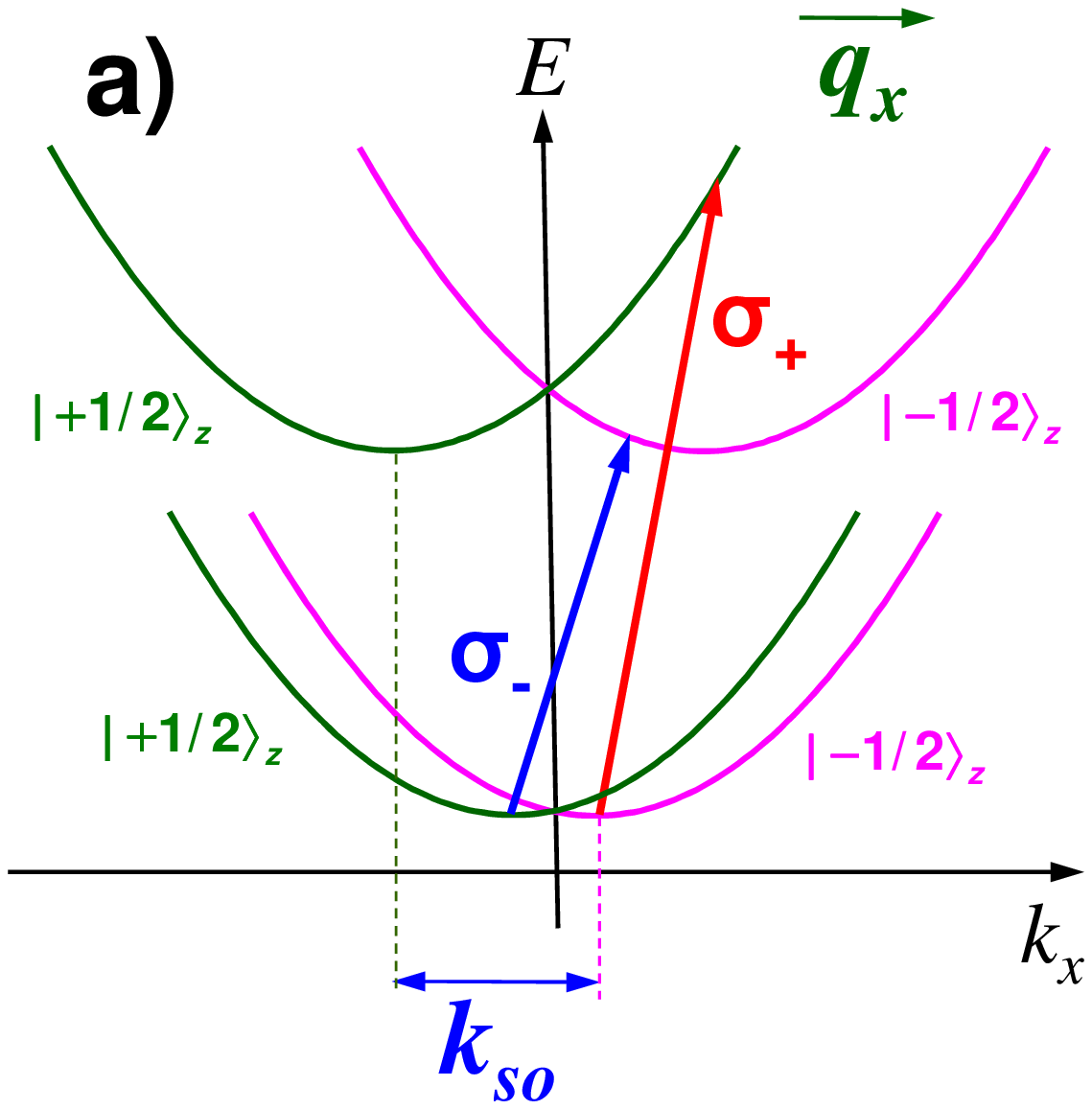} 
\quad
\includegraphics[width=0.3\linewidth]{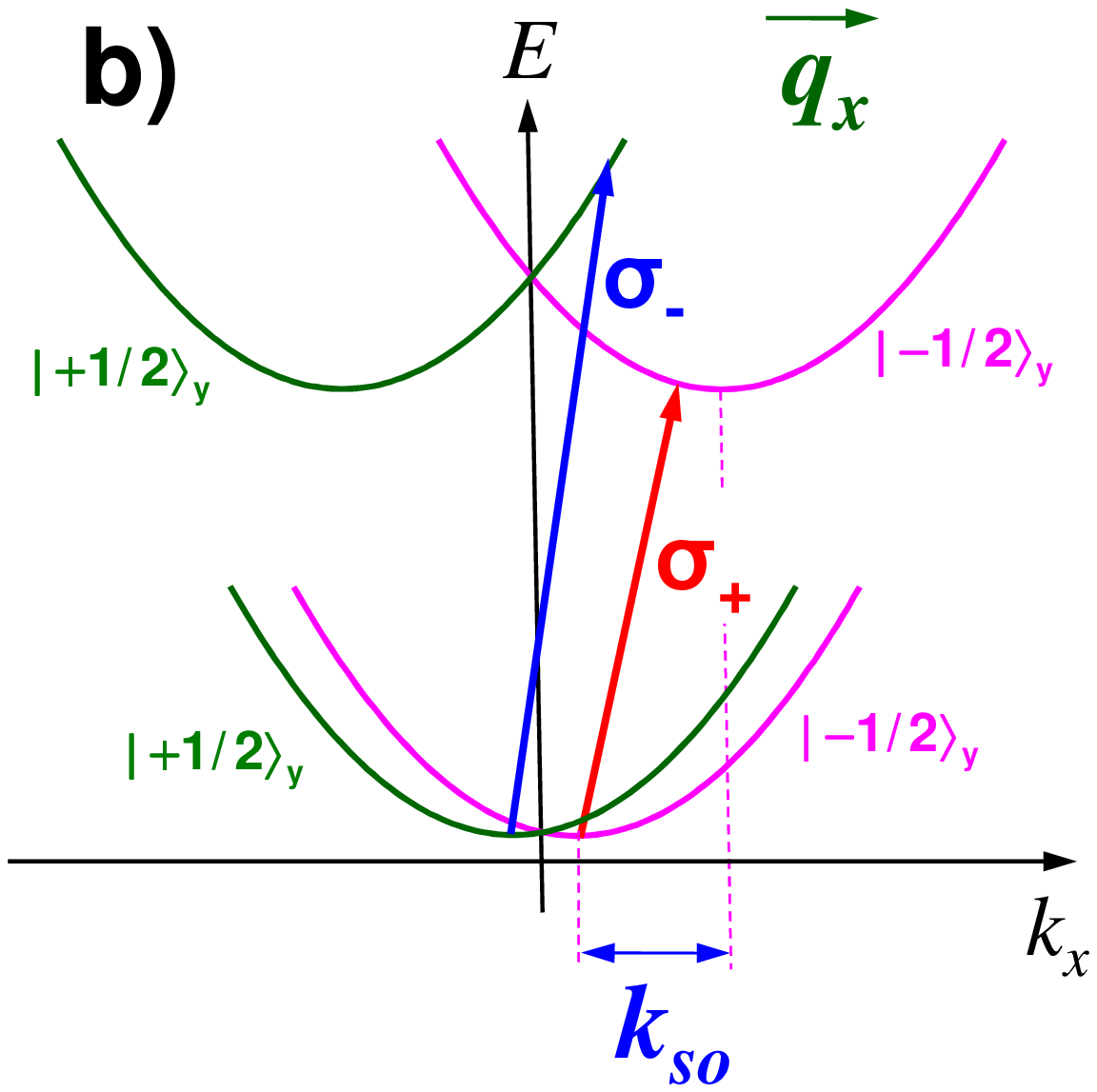}
\quad
\includegraphics[width=0.3\linewidth]{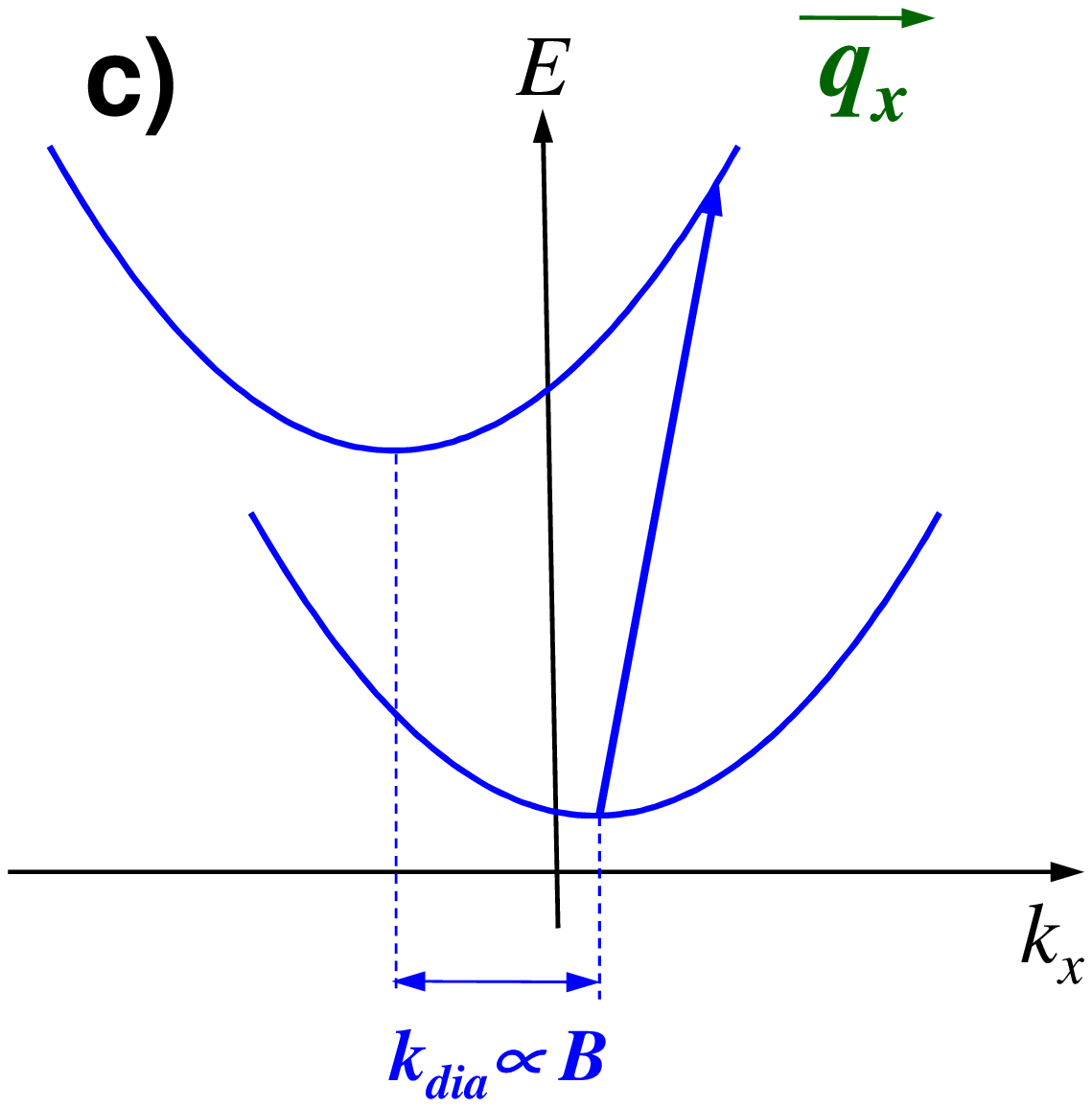}
\caption{Microscopic model of circular dichroism at intersubband transitions (a) in (110) and (311) QWs and (b) in (001) QWs. (c): Model of $\bm B$-dependent absorption.}\label{fig1}
\end{figure*}

If we include the spin-orbit interaction~\eqref{SO}, the energy dispersions in the subbands of size quantizations become splitted parabolas as it is shown in Fig.~\ref{fig1}a,b. 
For (110) and (311) QWs the term $\beta_{zx} \hat{\sigma}_z k_x$ is present in the spin-orbit Hamiltonian~\eqref{SO}, therefore the subbands for electrons with spin projection $\pm1/2$ onto the $z$ axis
are splitted. The selection rules for intersubband absorption of light propagating along $z$ direction yield higher transition rate from $\left|-{1/ 2}\right>_z$ state in the ground subband to the $\left|+{1/ 2}\right>_z$ state in the excited subband for $\sigma_+$ radiation, while for left-handed polarization the transition between two other states is more probable~\cite{ResInvCPGE}. This is illustrated in Fig.~\ref{fig1}a. 
As a result, we have in the energy conservation law~\eqref{energy} instead of the $q_x$ component
\begin{equation}
\label{kso}
	q_x \to q_x \mp n_z k_{so}.
\end{equation}
Here the upper and lower signs correspond to $\sigma_+$ and $\sigma_-$ polarizations, $\bm n = \bm q/q$ is a unit vector in the light propagation direction, and $k_{so}=(\beta_{zx}^{(1)}+\beta_{zx}^{(2)}){m/\hbar^2}$, where $\beta_{zx}^{(1,2)}$ are the corresponding spin-orbit constants for the first and the second subbands.

For (001) QWs the term $\beta_{yx} \hat{\sigma}_y k_x$ splits the states with a definite spin projection onto the $y$ axis. According to the selection rules, ``spin-conserving'' transitions between the states 
with the same spin projection are more probable for light with $\bm n \parallel y$ as it is shown in Fig.~\ref{fig1}b. Therefore the replacement~\eqref{kso} again takes place, but with $k_{so}=(\beta_{yx}^{(1)}-\beta_{yx}^{(2)}){m/\hbar^2}$ and with $n_y$ instead of $n_z$.

Substituting~\eqref{kso} into the $q_x^2$-contribution in Eq.~\eqref{eta_q}, we obtain the circular dichroism of the intersubband absorption:
\begin{equation}
\label{Delta_eta}
	\eta_{\sigma_+} - \eta_{\sigma_-} \propto  k_{so} q_x n_{z,y} \left[{\partial\over\partial (\hbar\omega)} - \overline{E} \: {\partial^2\over\partial (\hbar\omega)^2}\right]\eta_0(\hbar \omega - E_{21}),
\end{equation}
where the factor $n_z$ should be taken for (110), (311) QWs, and $n_y$ is for (001) QWs.
Angular dependence of the circular dichroism~\eqref{Delta_eta} 
coincides with Eqs.~\eqref{NOA_D2d} and~\eqref{NOA_Cs} obtained from symmetry arguments.

In the presence of a magnetic field, the diamagnetic interaction Eq.~\eqref{H_dia} results in spin-independent shift of the second subband relative to the first one making the energy separation $\bm k$-dependent:
\begin{equation}
	E_{21}(\bm k) = E_{21} + \alpha_{\mu\nu}B_\mu k_\nu.
\end{equation}
Therefore, similar to Eq.~\eqref{kso} we have
\[
q_x \to q_x + k_{dia},
\]
where $k_{dia}=\alpha_{\mu x}B_\mu \, m/\hbar^2$, Fig.~\ref{fig1}c. 
For (110) and (311)~QWs $\alpha_{z x}\neq 0$ so that $k_{dia} \propto B_z$, while for (001)~QWs $\alpha_{y x}\neq 0$ and $k_{dia} \propto B_y$.
As a result, we obtain the contribution to the absorbance
\begin{equation}
\label{Delta_eta_B}
	\Delta\eta(\bm B) \propto  q_x B_{z,y}  \left[{\partial\over\partial (\hbar\omega)} - \overline{E} \: {\partial^2\over\partial (\hbar\omega)^2}\right]\eta_0(\hbar \omega - E_{21}).
\end{equation}
This expression corresponds to phenomenological Eqs.~\eqref{MSD_C_2v} and~\eqref{MSD_110}.

\textbf{4. Theory.}
\begin{figure*}[tb]
\includegraphics[width=0.4\linewidth]{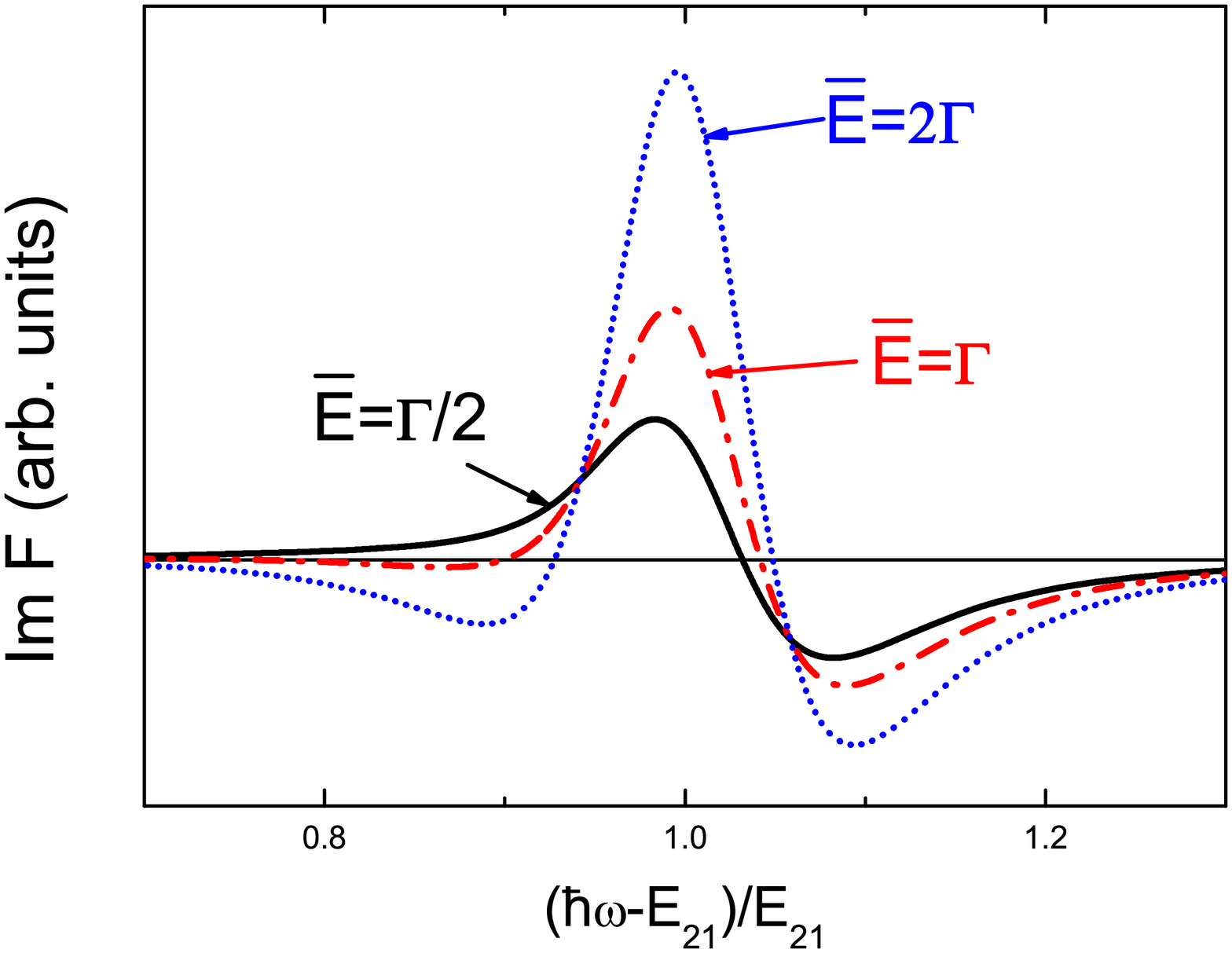}
\qquad
\includegraphics[width=0.4\linewidth]{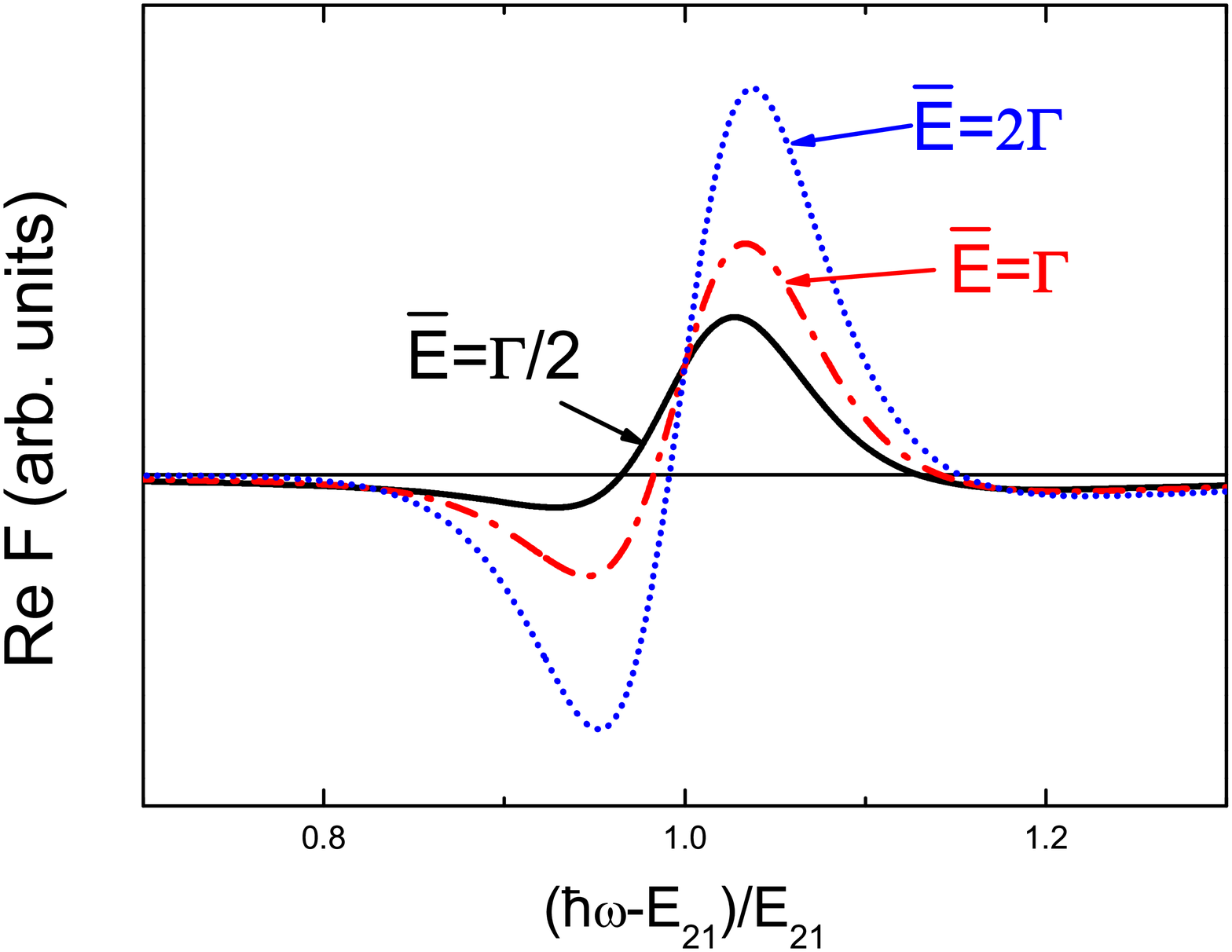}
\caption{Spectral dependences of gyrotropy and magneto-spatial dispersion constants 
for various mean electron energies $\overline{E}$. The intersubband absorption peak halfwidth is taken $\Gamma=0.1 E_{21}$.}
\label{fig2}
\end{figure*}
Now we develop a rigorous expression for gyrotropy and magneto-spatial dispersion constants in MQW structures.
Resonant contribution to the gyrotropy tensor has the form~\cite{IP74}
\begin{equation}
\label{gamma_def}
	\gamma_{\lambda\mu\nu}= 
	{4\pi{\rm i} \over\omega^2  d} \lim_{\bm q \to 0} {\partial\over\partial q_\nu} \sum_{\bm k,s,l} {j^\lambda_{1s,2l} \, j^\mu_{2l,1s} \: f_s(\bm k) \over E_{2l}(\bm k + \bm q) - E_{1s}(\bm k)  -\hbar\omega  -{\rm i}\Gamma}.
\end{equation}
Here $d$ is a period of MQW structure, the indices 
$s,l$ enumerate eigenstates in the ground and the excited subbands found with account for the spin-orbit interaction, $\hat{\bm j}$ is the electric current operator, $\Gamma$ is the halfwidth of the absorption peak, and $f_s(\bm k) = f_0(E_{1s}(\bm k))$, where $f_0(E)$ is the Fermi-Dirac distribution function. In the basis of these states
\[
E_{1s}(\bm k) = E_k + H^{(1)}_{ss}(\bm k),
\:
E_{2l}(\bm k) = E_{21} + E_k + H^{(2)}_{ll}(\bm k),
\]
where $\hat{H}^{(1,2)}(\bm k)$ are operators of the spin-orbit interaction~\eqref{SO} for the first and the second subbands, and $E_k=\hbar^2k^2/(2m)$.
In Eq.~\eqref{gamma_def} we assume that the electric current operators for intersubband transitions are independent of wavevector.

Expanding Eq.~\eqref{gamma_def} to the linear order in the spin-orbit interaction and performing summation over $\bm k$ we obtain
\begin{eqnarray}
	\label{gamma_fin}
	\gamma_{\lambda\mu\nu} = -{{\rm i}\over 2} {\rm Tr} \left[ \hat{L}^{\dag}_\lambda  {\partial \hat{H}^{(2)} \over \partial k_\nu} \hat{L}_{\mu}  - \hat{L}_\mu {\partial \hat{H}^{(1)} \over \partial k_\nu} \hat{L}^{\dag}_\lambda \right] \, F(\omega).
\end{eqnarray}
Here $\hat{\bm L}$ is defined according to $\hat{\bm j}=	\hat{\bm L}ev^z$, where $v^z = -({\rm i}\hbar/m)\partial/\partial z$, 
\begin{equation}
\label{F}
	F(\omega)= 
	\left[{\partial\over\partial (\hbar\omega)} - \overline{E}{\partial^2\over\partial (\hbar\omega)^2} \right]
	\epsilon_{21}(\hbar\omega),
\end{equation} 
the resonant contribution to the dielectric permittivity is given by
\begin{equation}
\label{eps_21}
\epsilon_{21}(\hbar\omega) = {4\pi N \over\omega^2 d} 
{e^2 |v^z_{21}|^2 \over E_{21} -\hbar\omega -{\rm i}\Gamma},
\end{equation}
$N$ is the 2D electron concentration, and the mean energy is defined as
$\overline{E} = 2\sum_{\bm k} E_k f_0(E_k)/ N$.

The operators $\hat{L}_\lambda$ for intersubband transitions have the form~\cite{ResInvCPGE}
\begin{equation}
\label{L}
\hat{\bm L}=({\rm i}\Lambda\hat{\sigma}_y,-{\rm i}\Lambda\hat{\sigma}_x,\hat{I}).
\end{equation}
Here $\hat{I}$ is the $2\times 2$ unit matrix, and
\[
\Lambda = {E_{21}\Delta(2E_g+\Delta) \over 2E_g (E_g+\Delta) (3E_g+2\Delta)} ,
\]
where  $E_g$ and $\Delta$ are the energy gap and the spin-orbit splitting of the valence band in the bulk semiconductor, respectively.

For (001) QWs Eqs.~\eqref{gamma_fin} and~\eqref{L} yield
\begin{equation}
\label{gamma}
	\gamma=\Lambda (\beta_{yx}^{(2)}-\beta_{yx}^{(1)})
	F(\omega),
\quad
\gamma'=\Lambda (\beta_{xy}^{(2)}-\beta_{xy}^{(1)})F(\omega),
\end{equation}
and for (110) and (311) QWs we obtain
\begin{equation}
\label{gamma1}
	\gamma_1 = \Lambda^2 (\beta_{zx}^{(2)}+\beta_{zx}^{(1)}) F(\omega).
\end{equation}
The remained constant $\gamma_2 \ll \gamma_1$ because the two-dimensional electrons do not feel the normal component of the photon momentum $q_z$ which is much smaller than the inverse MQW structure period $d^{-1}$.

Equation~\eqref{gamma_def} is also valid for calculation of the value $-{\rm i}A_{\lambda\mu\rho\nu}B_\rho$. Taking into account the diamagnetic interaction Eq.~\eqref{H_dia}, we obtain
\begin{equation}
	A_{zz\mu\nu} = \alpha_{\mu\nu} \, F(\omega).
\end{equation}
For (001) QWs this leads to Eq.~\eqref{MSD_C_2v} with
\begin{equation}
\label{A_1_2}
	A_1 = \alpha_{xy} \, F(\omega),
	\qquad
	A_2 = \alpha_{yx} \, F(\omega).
\end{equation}
For (110) and (311) QWs we obtain Eq.~\eqref{MSD_110} where 
\begin{equation}
\label{A_3}
	A_3 = \alpha_{zx} \, F(\omega),
\end{equation}
and $A_4 \ll A_3$.

\textbf{5. Discussion.}
Equations~\eqref{gamma},~\eqref{gamma1} and~\eqref{A_1_2},~\eqref{A_3}  demonstrate that the spectral dependence of the gyrotropy constants $\gamma$, $\gamma'$  and $\gamma_1$ as well as of magneto-spatial dispersion constants $A_{1..3}$ is given by the function $F(\omega)$, Eq.~\eqref{F}. However, $\gamma$ and $\gamma'$ are determined by a difference of corresponding spin-splitting constants in the excited and ground subbands, while $\gamma_1$ is proportional to the sum $\beta_{zx}^{(2)}+\beta_{zx}^{(1)}$. All these findings agree with the microscopic model, cf. Eqs.~\eqref{F} and~\eqref{Delta_eta}. The magneto-spatial dispersion in all cases is given by $\alpha_{\mu\nu}$ which describes the shift of the second subband dispersion relative to the first one.
The dependences ${\rm Im} F(\omega)$ and ${\rm Re} F(\omega)$ presented in Fig.~\ref{fig2} demonstrate sharp spectral changes in the resonance region. The relation between the mean energy $\overline{E}$ and the absorption peak halfwidth $\Gamma$ governs asymmetry of the resulting spectra of gyrotropy and magneto-spatial dispersion constants.

Note that $\bm k$-linear spin-dependent terms are also possible in the optical transition operators~\cite{SST_ST_EL}. They yield an additional contribution to the gyrotropy tensor with the frequency dependence coinciding with $\epsilon_{21}(\hbar\omega)$, Eq.~\eqref{eps_21}.

In order to estimate a magnitude of the effect, we note that in real systems $\Gamma \sim \overline{E} \sim 10$~meV~\cite{ResInvCPGE}, therefore for (001) MQWs we have
\[
\delta\varepsilon \sim \Lambda {\beta q   \over \Gamma } \, \epsilon_{21},
\]
and for GaAs based QWs with $\beta q \sim 10^{-3}$~meV, $\Lambda \sim 10^{-2}$, $N = 10^{12}~\mbox{cm}^{-2}$,  $\hbar\omega \sim E_{21} \sim 100$~meV, the experimentally measured quantity $\delta\varepsilon \: d \: {\omega n/ c} \sim  10^{-7} \, {\cal N}$, where ${\cal N}$ is a number of QWs in the MQW structure. In multi-pass geometry this value may be increased by an order of magnitude. Such signals can be detected in experiments. The estimate for (110) and (311) QWs has an additional small factor $\Lambda$ which is partially compensated by a larger  spin-orbit splitting constant $\beta_{zx}$. The situation is more favourable in $p$-type structures where the selection rules for circular polarization yield $\Lambda \sim 1$. However,  intersubband absorption peak is broadened in $p$-type QWs due to complicated band structure of the valence band.

The magneto-spatial dispersion effects can be estimated as
\[
\delta\varepsilon^{(\bm B)} \sim {\alpha B q   \over \Gamma } \, \epsilon_{21}.
\]
The main contribution to the diamagnetic $\bm k$-linear shift in (001) QWs is given by $\alpha_{xy}=-\alpha_{yx}= (z_2-z_1){e\hbar\over mc}$, where $z_n$ is the coordinate matrix element in the $n$th subband. Estimation for $z_2-z_1=10$~\AA, $d=100$~\AA \, and $B=1$~T yields $\delta\varepsilon^{(\bm B)}  d \: {\omega n/ c} \sim  10^{-5} \, {\cal N}$.

\textbf{6. Conclusion.}
In conclusion, we have studied gyrotropic and magneto-spatial properties of MQW structures showing up at intersubband optical transitions. 
Phenomenological model and microscopic theory are developed for MQWs of various crystallographic orientation. 
Gyrotropy is shown to be caused by spin-orbit splitting of size-quantized subbands while magneto-spatial dispersion arises from diamagnetic $\bm k$-linear shift of the intersubband energy gap.
It is demonstrated that the spectral dependence of the gyrotropy and magneto-spatial dispersion constants represents an asymmetrical peak with a degree of asymmetry governed by the mean electron energy. The estimates show that the considered effects are within the experimental sensitivity.

\acknowledgments{Author thanks S.A. Tarasenko for helpful discussions. This work was partially supported by RFBR and President grant for young scientists.

\end{document}